\documentclass[twocolumn]{aastex62}

\usepackage{amsmath}
\usepackage{appendix}

\def\xeff{\ensuremath\chi_\mathrm{eff}}
\def\xeffj{\ensuremath\chi_{\mathrm{eff},j}}
\def\postdist{\ensuremath p \big( \, \mu, \sigma^2 \,|\, \{ d_i \}^N_{i=1} \, \big)}
\def\xp{\ensuremath\chi_p}

\graphicspath{{./}{figures/}}
\shortauthors{Miller et al.}

\begin{document}

\title{The Low Effective Spin of Binary Black Holes and Implications for Individual Gravitational-Wave Events}

\correspondingauthor{S. Miller}
\email{smiller07@smith.edu}

\author{Simona Miller}
\affiliation{Department of Physics, Smith College, Northampton, MA 01063, USA}

\author{Thomas A. Callister}
\affiliation{Center for Computational Astrophysics, Flatiron Institute, New York, NY 10010, USA}

\author{Will M. Farr}
\affiliation{Center for Computational Astrophysics, Flatiron Institute, New York, NY 10010, USA}
\affiliation{Department of Physics and Astronomy, Stony Brook University, Stony Brook NY 11794, USA}

\begin{abstract}
While the Advanced LIGO and Virgo gravitational-wave experiments now regularly observe binary black hole mergers, the evolutionary origin of these events remains a mystery.
Analysis of the binary black hole spin distribution may shed light on this mystery, offering a means of discriminating between different binary formation channels.
Using the data from Advanced LIGO and Virgo's first and second observing runs, here we seek to carefully characterize the distribution of effective spin $\xeff$ among binary black holes, hierarchically measuring the distribution's mean $\mu$ and variance $\sigma^2$ while accounting for selection effects and degeneracies between spin and other black hole parameters.
We demonstrate that the known population of binary black holes have spins that are both small, with $\mu\approx0$, and very narrowly distributed, with $\sigma^2\leq0.07$ at 95\% credibility.
We then explore what these ensemble properties imply about the spins of \textit{individual} binary black hole mergers, re-analyzing existing gravitational-wave events with a population-informed prior on their effective spin.
Under this analysis, the binary black hole GW170729, which previously excluded $\xeff=0$, is now consistent with zero effective spin at $\sim10\%$ credibility.
More broadly, we find that uninformative spin priors generally yield \textit{overestimates} for the effective spin magnitudes of compact binary mergers.
\vspace{1cm}
\end{abstract}

\section{Introduction}
\label{sec:intro}


In their first two observing runs, the Advanced LIGO (Laser Interferometer Gravitational-wave Observatory) and Advanced Virgo experiments~\citep{aLIGO,aVirgo} have detected gravitational waves from eleven compact binary coalescences - ten binary black hole (BBH) mergers and one binary neutron star (BNS) merger~\citep{O1-BBH,GW170817,O2-Catalog}.
This early catalog of gravitational-wave (GW) signals has already been used to measure a diverse set of physical parameters, from the binaries' component masses and spins~\citep{GW150914-PE,GW170817-PE,O2-Catalog,Chatziioannou2019,Kimball2019} to more complex phenomena such as tidal effects~\citep{GW170817-EOS,Raithel2018} and consistency with general relativity~\citep{GW150914-testingGR,GW170817-testingGR,O2-TestingGR}.

\vspace{3px}

The number of gravitational-wave detections is rapidly growing.
Now roughly halfway through their third ``O3'' observing run, LIGO and Virgo have already reported $\sim40$ new detection candidates; together with the newly-built KAGRA detector~\citep{KAGRA}, they are expected to observe an additional $\sim100$ events in O4~\citep{O4_prospects}.
As the number of gravitational-wave detections increases, we can shift our focus from the study of individual binaries to the analysis of their ensemble~\citep{O1-BBH,O2-Populations}.
While the astrophysical details of compact binary evolution are only weakly reflected in the properties of individual events, we expect them to much more strongly inform binaries' population properties -- the distributions of component masses, spins, and redshifts.
Uncovering these ensemble properties is therefore a promising means of determining which evolutionary channel drives the formation of compact binary mergers: the evolution of isolated stellar binaries, dynamical capture in dense stellar environments, or yet another channel altogether.

\vspace{3px}

In this paper, we will explore the distribution of binary black hole spins detected by Advanced LIGO and Virgo.
Ours is not the first attempt at such an analysis -- several authors have previously explored the binary black hole spin distribution, using varying subsets of the LIGO and Virgo detections and employing a range of different models.
Our goals in this work are threefold:

\textit{1.~Focus on the observable}.
Many previous studies aim to describe the distributions of spin magnitudes and/or tilt angles of the individual black holes comprising the observed BBH population~\citep{Talbot2017,Farr2017a,Farr2018,Tiwari2018,Fernandez2019,Wysocki2019, Stevenson2017,O2-Populations,Roulet2019}.
Gravitational waves carry relatively little information, however, about the spins of these individual components.
At leading order, gravitational-wave signals instead depend only on a single effective spin parameter $\xeff$, quantifying the net projection of both component spins onto a binary's orbital angular momentum~\citep{Damour2001,Racine2008,Ajith2011,Ng2018,Roulet2019}.
We will therefore not attempt to model the ensemble of underlying component spins, but instead seek to describe the more readily-measurable $\xeff$ distribution \citep{Farr2017a,Farr2018,Tiwari2018,Roulet2019}.

\textit{2.~Adopt a simple population model.}
Popular models for the binary black hole spin distribution are often rather complex.
The flagship LIGO/Virgo analysis, for example, employs a flexible five-parameter model to describe the ensemble of black hole spins~\citep{O2-Populations}.
With only ten detections information these measurements, the resulting constraints from such models are generally uninformative.
Here, we will instead adopt a deliberately simple model, characterizing the $\xeff$ distribution via only two parameters: its mean and variance.
While this choice sacrifices some flexibility, it will yield transparent and intuitive results that are more robustly extracted from presently small number of gravitational-wave events.

\textit{3.~Account for observational biases}.
The underlying $\xeff$ distribution is observationally obscured in two ways.
The first is selection bias -- more highly spinning binary systems accumulate more gravitational-wave cycles, and are thus preferentially detected with higher signal-to-noise ratios.
The second is potential measurement degeneracy between effective spin and other binary parameters, most notably the mass ratio~\citep{Ng2018}.
Previous analyses variously do~\citep{Tiwari2018,Roulet2019,O2-Populations} and do not~\citep{Farr2017a,Farr2018,Fernandez2019} account for these spin-dependent effects.

\vspace{3px}

Just as individual binary black holes inform measurements of the population's spin distribution, so too can knowledge of the spin distribution inform our conclusions about individual events.
After hierarchically measuring the binary black hole $\xeff$ distribution (Sec.~\ref{sec:results}), we will therefore turn around and use this information to re-analyze the ten LIGO/Virgo detections, producing updated measurements of their effective spins (Sec.~\ref{sec:refined_measurements}).

\section{Hierarchical Inference of Black Hole Spins}
\label{sec:inference}

At leading order, the amplitude and phase evolution of a binary's gravitational-wave signal depend on its component spin through two parameters: the ``effective'' and ``precessing'' spins, $\xeff$ and $\xp$~\citep{Ajith2011,Schmidt2015}.
The effective spin $\xeff$, a unitless quantity between $-1$ and $1$, is the mass-weighted average of the dimensionless component spins $\vec a_1$ and $\vec a_2$, projected along the unit vector $\hat L_N$ parallel to the system's orbital angular momentum:
    \begin{equation}\label{eqn:Xeff1}
    \xeff = \frac{\left( m_1\,\vec a_1 + m_2\,\vec a_2 \right) \cdot \hat{L}_N}{m_1 + m_2}.
    \end{equation}
Here, $m_i$ are the source-frame masses of each black hole such that $m_1 > m_2$.
The effective spin may also be expressed in terms of the component spin magnitudes $a_i$ and tilt angles $t_i$ between each spin and $\hat L_N$:
\begin{equation}\label{eqn:Xeff2}
    \xeff = \tfrac{1}{m_1 + m_2} \left( a_1m_1\cos t_1 + a_2m_2\cos t_2 \right).
\end{equation}
$\xp$, on the other hand, is a unitless value between 0 and 1 that parametrizes the leading-order effects of orbital precession due to misaligned spins.
While $\xeff$ is related to the spin components parallel to $\hat L_N$,  $\xp$ is a linear combination of spin components \textit{normal} to $\hat L_N$.

\vspace{3px}

In this paper, we will focus on understanding the distribution of $\xeff$ across the population of merging stellar-mass black holes.
In contrast to more complex, many-parameter models featured in previous work~\citep{O2-Populations,Wysocki2019}, we will assume that effective spins are drawn from a simple truncated Gaussian,
    \begin{equation} \label{eqn:gaussian}
    p(\xeff | \mu, \sigma^2) =  \mathcal{N}(\mu,\sigma^2) \, \mathrm{exp}\bigg[ \frac{-(\xeff - \mu)^2}{2 \sigma^2}\bigg],
    \end{equation}
and to seek to measure the mean $\mu$ and variance $\sigma^2$ of this ensemble distribution.
The normalization constant
\begin{equation}\label{eqn:gaussian_normalization}
    \mathcal{N}(\mu,\sigma^2) = \sqrt{\frac{2}{\pi \sigma^2}} \,\, \Bigg( \mathrm{erf}\bigg[ \frac{1-\mu}{\sqrt{2 \sigma^2}}\bigg] + \mathrm{erf}\bigg[ \frac{1+\mu}{\sqrt{2 \sigma^2}}\bigg] \Bigg)^{-1}.
\end{equation}
ensures that Eq.~\eqref{eqn:gaussian} is properly normalized over the range $\xeff\in[-1,1]$.

\vspace{3px}

Although $\xeff$ is measured far more precisely than the individual component spins, it is still subject to significant uncertainty~\citep{Ng2018,O2-Catalog}.
More specifically, for each gravitational-wave signal (with associated data $d$), we do not obtain a direct point estimate of $\xeff$, but instead a set of discrete samples from the posterior probability distribution $p(\xeff|d)$ for the effective spin.
We must therefore employ a hierarchical approach in which each LIGO/Virgo BBH event is assumed to have some true but unknown value of $\xeff$ drawn from Eq.~\eqref{eqn:gaussian} above.
We will marginalize over the possible $\xeff$ of each LIGO/Virgo event to obtain a posterior on the mean $\mu$ and variance $\sigma^2$ of the population's effective spin distribution.

Given $N$ independent binary black hole detections (in our case $N=10$) with population-averaged detection efficiency $\xi$ and data $\{d_i\}_{i=1}^N$, the resulting posterior distribution for $\mu$ and $\sigma^2$ is given in Eq.~\eqref{eqn:posterior}~\citep{Loredo1995,Loredo2004,Fishbach2018,Mandel2019}.
The first line corresponds to the ideal case in which the posterior $p(\xeff,m_1,m_2,z|d_i)$ on each event's effective spin, component masses, and redshift $z$ are exactly known; the second line is for the realistic scenario in which we have only discrete samples from the posterior of every event.
    \begin{widetext}
    \begin{equation}
    \begin{aligned}
    \label{eqn:posterior}
        \postdist &\propto
            \frac{p(\mu,\sigma^2)}{\xi(\mu, \sigma^2)^N} \, \prod^N_{i=1}
            \left[\,
                \int d\chi_\mathrm{eff}\,dm_1\, dm_2\, dz\,\,
                    p(\xeff,m_1,m_2,z \,|\, d_i)
                    \frac{p_\mathrm{astro}(m_1,m_2,z)}{p_\mathrm{pe}(m_1, m_2, z)}
                    \frac{p(\xeff\,|\,\mu,\sigma^2)}{p_\mathrm{pe}(\xeff)}
            \right] \\
        & \propto \frac{p(\mu,\sigma^2)}{\xi(\mu, \sigma^2)^N} \, \prod^N_{i=1}
            \bigg\langle
                \frac{p_\mathrm{astro}(m_{1,j},m_{2,j},z_j)}{p_\mathrm{pe}(m_{1,j}, m_{2,j}, z_j)}
                \frac{p(\chi_{\mathrm{eff},j} \,|\, \mu, \sigma^2)}{p_\mathrm{pe}(\chi_{\mathrm{eff},j})}
            \bigg\rangle_{\mathrm{samples}\,j}
    \end{aligned}
    \end{equation}
    \end{widetext}

The effective spin samples generated by parameter estimation are necessarily obtained under the assumption of some default prior $p_\mathrm{pe}(\xeff)$.
The LIGO/Virgo samples produced by the \texttt{LALInference} software~\citep{Veitch2015,lalsuite}, for instance, are given by priors that are uniform in component spin magnitude and orientation, such that $p_\mathrm{pe}(a)$ and $p_\mathrm{pe}(\cos t)$ are both constant.
This corresponds to a symmetric $\xeff$ prior that is broady peaked about zero.
In Eq.~\eqref{eqn:posterior} we must ``undo'' this default prior and reweight each sample by the proposed Gaussian distribution $p(\chi_{\mathrm{eff},j} \,|\, \mu, \sigma^2)$, evaluated at the given sample value $\chi_{\mathrm{eff},j}$).

\vspace{3px}

\texttt{LALInference} also imposes a computationally-simple but unphysical prior $p_\mathrm{pe}(m_1,m_2,z)$ on the component masses and redshift of a binary black hole merger.
Specifically, uniform priors are adopted for the \textit{detector} frame masses $m_1(1+z)$ and $m_2(1+z)$ as well as the luminosity distance $D_L$.
This translates into~\citep{O2-Populations}
   \begin{multline}
    \label{eqn:pLAL}
    p_\mathrm{pe}(m_1,m_2,z) \\ \propto (1+z)^2 D_L(z)^2 \bigg[ D_c(z) +  \frac{c\,(1+z)}{H(z)}\bigg].
    \end{multline}
Here, $D_c(z)$ is the comoving distance at redshift $z$ and $c$ is the speed of light.
$H(z) =  H_0 \sqrt{\Omega_M (1+z)^3 + \Omega_\Lambda}$ is the Hubble parameter, given by the present-day Hubble constant $H_0= 67.27$ km/s/Mpc and energy densities $\Omega_M=0.3156$ and $\Omega_\Lambda= 0.6844$ of mass and dark energy, respectively~\citep{Planck2016}.
Although total mass $m_1+m_2$ is not strongly correlated with $\xeff$, $\xeff$ is generally \textit{anti-correlated} with a binary's mass ratio $q=m_2/m_1$, as shown in \cite{Roulet2019}, for example.
The default \texttt{LALInference} priors are uniform in detector-frame component masses, and thus preferentially tolerate systems with unequal mass ratios, thereby pushing $\xeff$ posteriors to larger values~\citep{Ng2018,Tiwari2018}.
We compensate for this potential bias in Eq.~\eqref{eqn:posterior} by further reweighting \texttt{LALInference} samples by an astrophysically-motivated mass and redshift prior, consistent with the measured mass and redshift distributions of binary black holes following the second Advanced LIGO/Virgo observing run~\citep{O2-Populations}:
    \begin{equation}
    \label{eqn:pASTRO}
    p_\mathrm{astro}(m_1,m_2,z) \propto \frac{(1+z)^{1.7}}{m_1 (m_1 - M_\mathrm{min})} \frac{d V_c}{d z}.
    \end{equation}
This prior is logarithmically-uniform in primary mass and uniform in mass-ratio, and assumes a binary merger rate that follows star formation, growing as $(1+z)^{2.7}$ in the source frame~\citep{Madau2014}.
In our detector frame, the measured merger rate is redshifted to $\frac{R(z)}{(1+z)} = (1+z)^{1.7}$.
$\frac{dV_c}{dz}$ is the comoving volume per unit redshift, and we assume a minimum black hole mass $M_\mathrm{min}=5\,M_\odot$.
We do not impose a maximum mass cutoff.

\vspace{3px}

\begin{figure}
    \centering
    \includegraphics[width=0.45\textwidth]{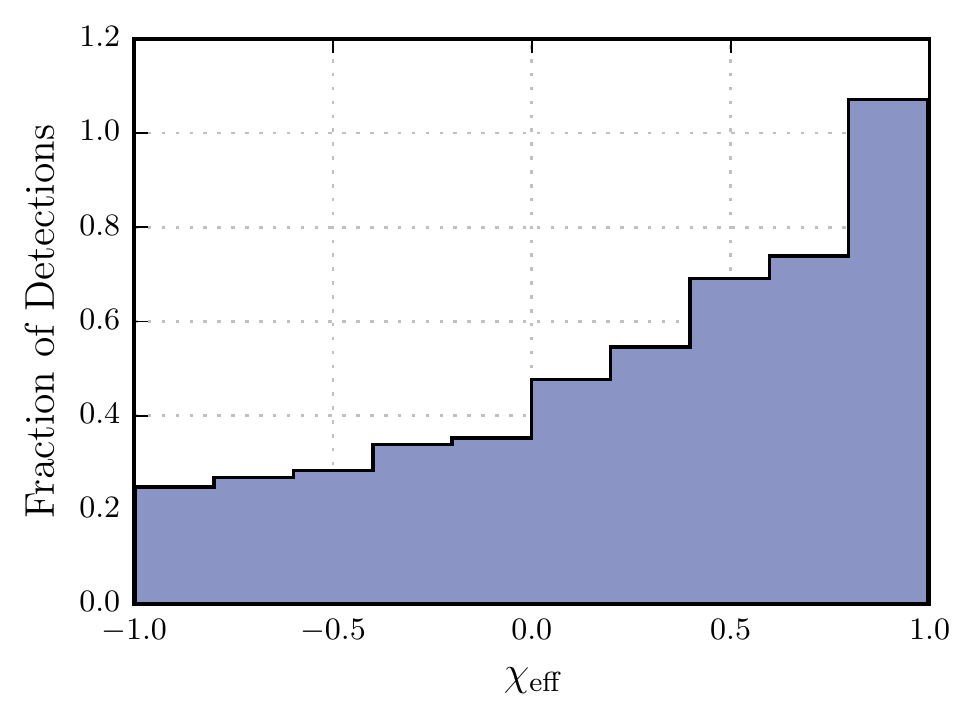}
    \caption{
    Relative numbers of ``detected'' synthetic BBHs as a function of $\xeff$, determined by selecting systems with single-detector matched filter signal-to-noise ratios $\rho\geq 8\sqrt{2}$.
    The synthetic events have random orientations and sky locations, masses and redshifts following Eq.~\eqref{eqn:pASTRO}, and effective spins drawn from a uniform distribution between $-1\leq \xeff\leq 1$.
    Detection probability increases monotonically with $\xeff$; at Advanced LIGO's current sensitivity, a binary with $\xeff=1$ is approximately five times as likely to be detected as one with $\xeff=-1$.
    This distribution of synthetic detections is used in Eq.~\eqref{eqn:selection-sum} to mediate the influence of selection effects on the measured $\xeff$ distribution.
    }
    \label{fig:selection}
\end{figure}

Observational selection effects are accounted for in Eq.~\eqref{eqn:posterior} by the population-averaged detection efficiency $\xi(\mu,\sigma^2)$, the fraction of all BBH mergers that LIGO successfully detects:
    \begin{equation}\label{eqn:xi}
    \xi(\mu, \sigma^2) = \int d\chi_\mathrm{eff}\, p(\xeff\, | \, \mu, \sigma^2)\, P_\mathrm{det}(\xeff)
    \end{equation}
where $P_\mathrm{det}(\xeff)$ is the probability, marginalized over all other parameters, that we will detect a BBH with a given $\xeff$.
In practice, we compute $\xi(\mu,\sigma^2)$ via the Monte Carlo approach of \cite{Farr2019}, drawing synthetic BBHs with random orientations, masses and redshifts following Eq.~\eqref{eqn:pASTRO}, and effective spins from a flat reference distribution $p_\mathrm{ref}(\xeff) \propto 1$.
We compute the Advanced LIGO matched filter signal-to-noise ratio $\rho$ for each synthetic BBH using the ``Early High-Sensitivity'' power spectral density of \cite{O4_prospects} and the precessing \texttt{IMRPhenomPv2} waveform model~\citep{IMRPhenomPv2}, although we assume purely aligned spins.
Events with $\rho\geq 8\sqrt{2}$ in the Advanced LIGO detector network are considered to have been ``detected.''
Figure~\ref{fig:selection} shows a histogram of the effective spins for these mock detections; at current sensitivities, we see that Advanced LIGO is roughly five times more likely to detect a maximally aligned system ($\xeff=1$) than a maximally anti-aligned system ($\xeff=-1$).
Given a proposed mean $\mu$ and variance $\sigma^2$ of the $\xeff$ distribution, the corresponding detection fraction $\xi(\mu,\sigma^2)$ can then be evaluated as a reweighted sum over our catalog of synthetic ``detections''~\citep{Farr2019},
    \begin{equation}
    \label{eqn:selection-sum}
    \xi(\mu,\sigma^2) = \frac{1}{N_\mathrm{draw}}\sum_{i=1}^{N_\mathrm{det}}
        \frac{p(\chi^\mathrm{det}_{\mathrm{eff},i} | \mu,\sigma^2)}{p_\mathrm{ref}(\chi^\mathrm{det}_{\mathrm{eff},i})},
    \end{equation}
where $N_\mathrm{draw}$ is the total number of synthetic events, $N_\mathrm{det}$ is the number of ``detections,'' and $\chi^\mathrm{det}_{\mathrm{eff},i}$ is the effective spin of the $i^\mathrm{th}$ detected event.

\vspace{3px}

Finally, we note that Eq.~\eqref{eqn:posterior} does not depend on the overall rate of black hole mergers.
As we are concerned only with the \textit{shape} of the $\xeff$ distribution and not the absolute \textit{number} distribution $dN/d\xeff$ of mergers, Eq.~\eqref{eqn:posterior} is derived by marginalizing over the total event rate, assuming a logarithmically uniform rate prior~\citep{Fishbach2018,Mandel2019}.

\section{Results from O1 and O2 Detections}
\label{sec:results}

\begin{figure*}
    \centering
    \includegraphics[width=0.7\textwidth]{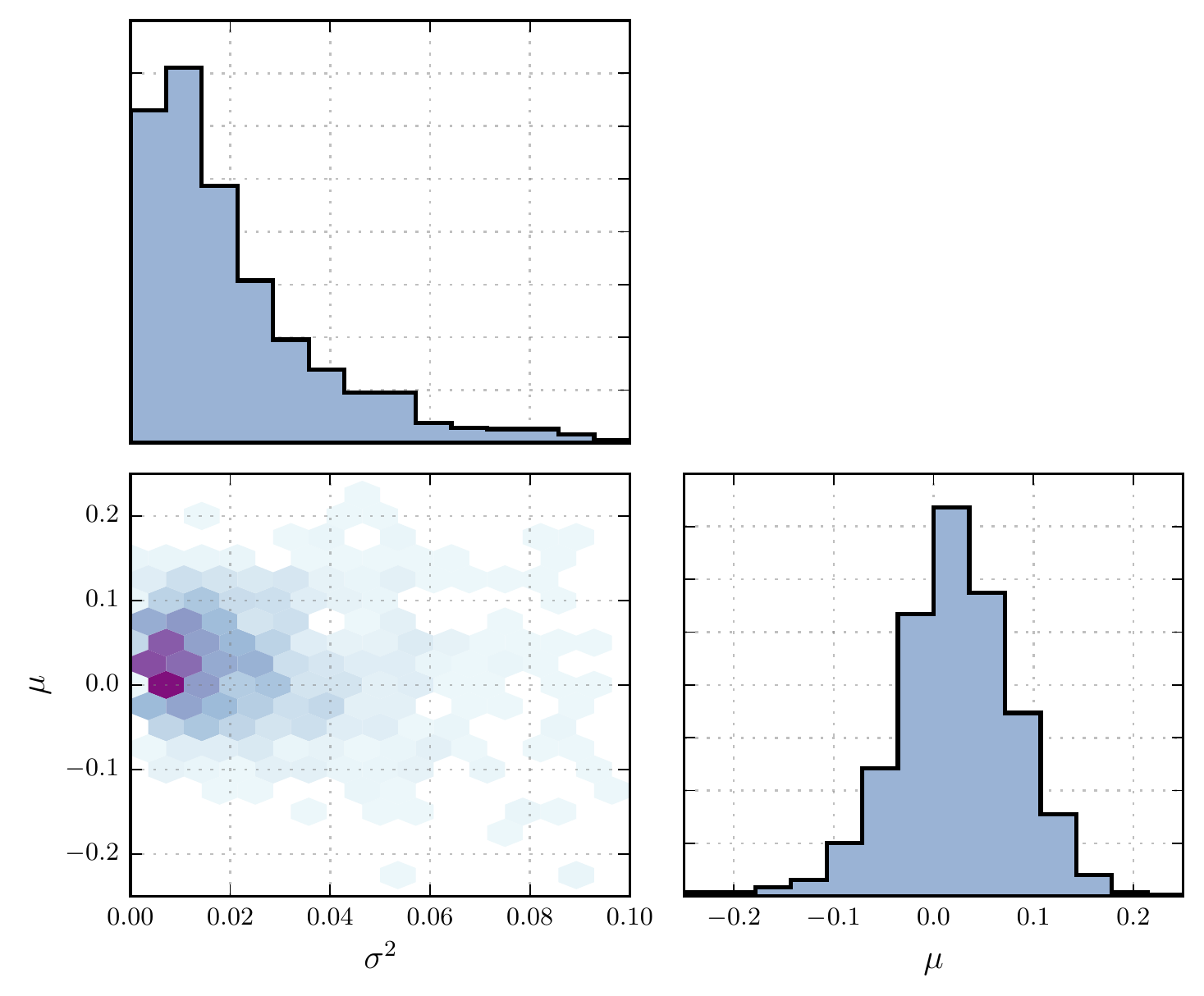}
    \caption{The posterior distributions on the mean $\mu$ and variance $\sigma^2$ of the $\xeff$ distribution of binary black hole mergers.
    Given the ten binary black holes observed by Advanced LIGO and Virgo in their first two observing runs, we find $\mu=0.02^{+0.11}_{-0.13}$ and $\sigma^2\leq 0.07$ at 95\% credibility.
    Notably, $\mu$ remains consistent with zero, as expected for binary black holes formed dynamically in dense stellar environments~\citep{Rodriguez2016,Rodriguez2018,Doctor2019,Rodriguez2019}, although isolated binary formation (which would predict $\mu>0$) remains plausible if black hole spins are intrinsically small~\citep{Farr2017a,Farr2018,Qin2018,Fuller2019}.
    Interestingly, and in contrast to previous results~\citep{Roulet2019}, we find that $\sigma^2$ is consistent with $0$; we therefore cannot rule out an arbitrarily narrow $\xeff$ distribution.
    }
    \label{fig:corner_plt}
\end{figure*}

\begin{figure}
    \centering
    \includegraphics[width=0.45\textwidth]{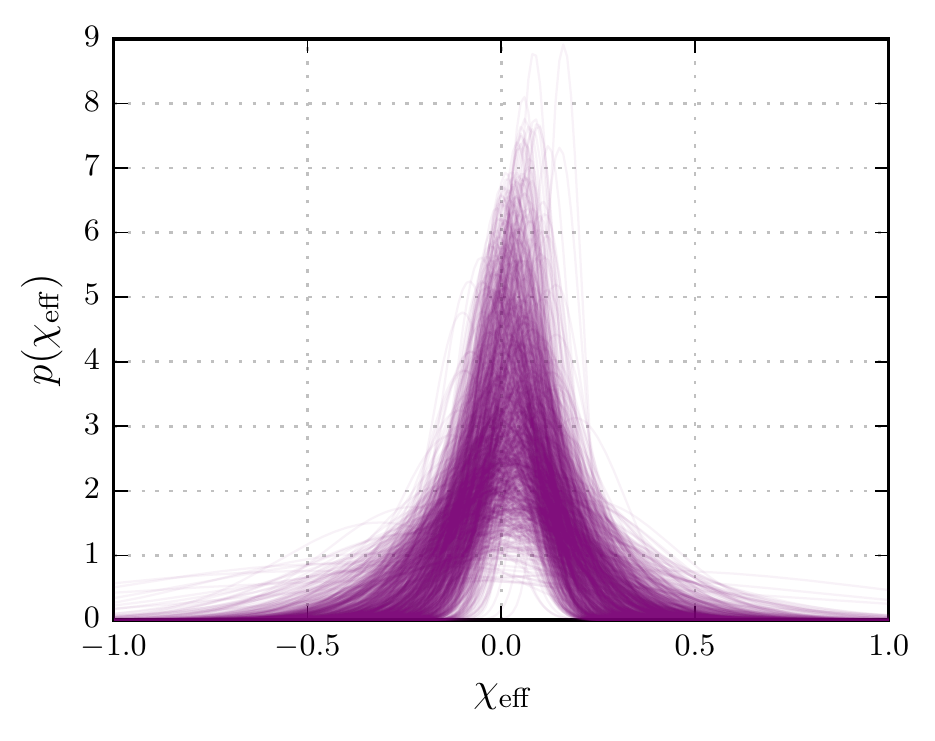}
    \caption{A superposition of the allowed Gaussian $\xeff$ distributions corresponding to the $(\mu,\sigma^2)$ samples shown in Fig.~\ref{fig:corner_plt}.
    The known population of binary black holes robustly requires a $\xeff$ distribution that is narrowly peaked about zero.
    We do, however, see a slight preference for positive $\xeff$, as previously shown in Fig.~\ref{fig:corner_plt}.}
\label{fig:prob_dist}
\end{figure}

We analyze the ten binary black hole mergers reported by LIGO and Virgo in their O1 and O2 observing runs~\citep{O2-Catalog, GWTC1_PE_samples}.
Before discussing results, it is useful to review expectations from the literature for the spin distributions resulting from different formation scenarios.
Isolated binary evolution is predicted to yield black hole with spins preferentially aligned with their orbit.
Although spin misalignments may be introduced by natal supernova kicks, episodes of mass transfer and tidal torques serve to realign component spins before the formation of the final black hole binary~\citep{Rodriguez2016,Zevin2017,Gerosa2018,Qin2018,Zaldarriaga2018,Bavera2019}.
The black holes' spin \textit{magnitudes} in this scenario are much more uncertain.
Recent work indicates that angular momentum is efficiently transported away from stellar cores, leaving black holes with natal spins as low as $a\sim10^{-2}$~\citep{Qin2018,Fuller2019}.
While tides on the progenitor of the second-born black hole can spin up the progenitor star~\citep{Zaldarriaga2018}, this effect can be counteracted by mass loss in stellar winds, and more detailed simulations find only low or moderate spin increases due to tides~\citep{Qin2018,Bavera2019}.
Meanwhile, dynamically-formed systems in dense stellar clusters have no \textit{a priori} preferred axis, and so are likely to have random spin configurations~\citep{Rodriguez2016,Rodriguez2018,Doctor2019,Rodriguez2019}.
Once again, however, the expected spin magnitudes are largely unknown, subject to the same uncertainties mentioned above regarding natal black hole spins.
One firm prediction of the dynamical scenario concerns the spins of \textit{second-generation} binaries, whose components were themselves formed from previous mergers.
Regardless of their component spins, black hole mergers generally yield remnants with $a\sim0.7$; thus the effective spin of two such second-generation binaries may be large~\citep{Fishbach2017,Gerosa2017,Rodriguez2018,Doctor2019,Rodriguez2019}.

\vspace{3px}

In summary, the most robust discriminator between the isolated binary and dynamical scenarios is the mean $\mu$ of the effective spin distribution.
If isolated binaries have preferentially aligned spins, then we expect an effective spin distribution centered on a positive value: $\mu>0$.
Dynamically-formed binaries with random spin orientations, meanwhile, should have a symmetric $\xeff$ distribution centered at $\mu = 0$~\citep{Farr2018}.

\vspace{3px}

To generate our posterior distributions for $\mu$ and $\sigma^2$, we sample from Eq.~\ref{eqn:posterior} using \texttt{emcee}, a Markov Chain Monte Carlo (MCMC) package in Python~\citep{emcee}.
We use the public \texttt{LALInference} samples made available through the Gravitational Wave Open Science Center~\citep{gwosc,GWTC1_PE_samples}.
Specifically, we use the so-called ``overall posterior" samples, which are a union of the samples obtained with the \texttt{IMRPhenomPv2}~\citep{IMRPhenomPv2} and \texttt{SEOBNRv3}~\citep{SEOBNRv3_1, SEOBNRv3_2} waveform models; we have confirmed that our results are robust under the use of either waveform model independently.
We adopt flat priors across the ranges $\mu\in[-1,1]$ and $\sigma^2 \in [0,1]$.
Our posterior on $\mu$ and $\sigma^2$ is given in Fig.~\ref{fig:corner_plt}.
With the ten binary black holes observed in O1 and O2 by Advanced LIGO and Virgo, we constrain the mean of the binary black hole $\xeff$ distribution to $\mu = 0.02^{+0.11}_{-0.13}$.
Hence we find no evidence for preferential spin alignment, which would manifest as preferentially positive $\mu$.
Meanwhile, we find that width of the $\xeff$ distribution must be extremely small, with a variance $\sigma^2<0.07$ at 95\% credibility.
Curiously, provided that $\mu$ is nonzero, the data are consistent with a vanishingly narrow spin distribution.
This can be seen in Fig.~\ref{fig:corner_plt}, which does not exclude $\sigma^2=0$.
We therefore cannot yet rule out the possibility that the $\xeff$ distribution is a delta function, with all binary black holes sharing the same effective spin value.

\vspace{3px}

In Fig.~\ref{fig:prob_dist}, we plot the ensemble of permitted $\xeff$ distributions consistent with our posterior on $\mu$ and $\sigma^2$.
This figure again shows our slight preference for positive $\mu$, but primarily illustrates that all $\xeff$ distributions consistent with the presently-known binary black hole mergers must be narrowly peaked about $\sim 0$.

\vspace{3px}

\begin{figure}
    \centering
    \includegraphics[width=0.45\textwidth]{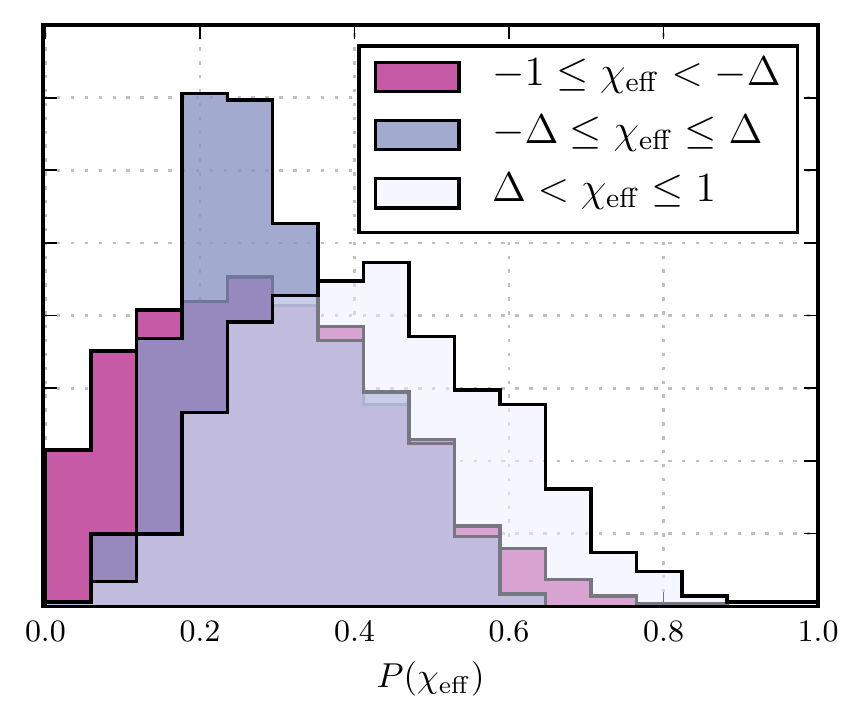}
    \caption{Histogram of the probability that there exists binary black holes with various effective spins $\xeff$.
    Here, $\Delta = 0.05$, representing a minimal detectable $\xeff$ value \citep{Farr2018, O2-Populations}.
    The data represents the area under each distribution plotted in Fig.~\ref{fig:prob_dist}, for each $\xeff$ range indicated.
    If the ten LIGO/Virgo binary black hole detections are a representative draw from the local universe's BBH population, then these results indicate that in future detections, positive $\xeff$ is more likely to be observed than negative $\xeff$.
    Notice that the negative $\xeff$ histogram includes $0$ while the other two do not.}
    \label{fig:chi_prob_ranges}
\end{figure}

Such small effective spins likely indicate one of three possibilities.
First, these results are compatible with a scenario in which component spins are large but preferentially lie in the plane perpendicular to the binary's orbital angular momentum.
Intriguingly, some models for binary mergers driven by Kozai-Lidov resonances~\citep{Kozai1962,Lidov1962} in hierachical triples predict preferentially perpendicular spins~\citep{Rodriguez2018_triple,Antonini2018,Liu2018},
although this effect is not observed in all studies~\citep{Antonini2018,Liu2019}.
Second, all component spins could be large but primarily anti-aligned with one another, such that each spin cancels the other's contribution to Eq.~\eqref{eqn:Xeff1}; we are not aware of any merger channels that predict this.
(Note, however, that BBH formation in the disks of active galactic nuclei may lead to the unique possibility of anti-aligned spins in approximately \textit{half} of the BBH population~\citep{McKernan2019,Yang2019}).
Finally, as has been noted by other authors, these results may point to the simple fact that binary black hole component spins are intrinsically small~\citep{Farr2017a,Tiwari2018,Wysocki2019}.
This final scenario would yield a $\xeff$ distribution concentrated around zero, regardless of spin orientation. By positing small natal spins, models of isolated field binaries~\citep{Belczynski2017,Postnov2019}, dynamical formation~\citep{Rodriguez2019}, and hierarchical triples~\citep{Antonini2018} can all produce distributions like those shown in Fig.~\ref{fig:prob_dist}.

Given the theoretical uncertainties in spin magnitude, some authors have suggested that, rather than $\mu$ and $\sigma^2$, a more robust prediction might be the \textit{fraction} of black hole binaries with negative $\xeff$~\citep{Rodriguez2016,Gerosa2018}.
In Fig.~\ref{fig:chi_prob_ranges}, we show posteriors on the fractions of binaries with positive and negative effective spins.
We follow \cite{Farr2018} and \cite{O2-Populations} in additionally defining a third ``uninformative'' bin $-\Delta\leq\xeff\leq\Delta$, with $\Delta = 0.05$, in which effective spins are essentially indistinguishable from zero.
At 95\% credibility, a fraction $0.28^{+0.26}_{-0.17}$ of binaries are expected to have uninformatively small effective spins.
Of the informative events, $0.41^{+0.38}_{-0.37}$ are predicted to have negative spins.
Notably, however, the posteriors in Fig.~\ref{fig:chi_prob_ranges} do not exclude the possibility of \textit{no} detectably-negative effective spins; instead we can only confidently limit the fraction of detectably negative effective spins to $\leq0.75$ at 95\% credibility.
These results differ from \cite{O2-Populations}, which reports that $\sim$80\% of BBH have uninformatively small spins.
This difference can be attributed to two possibilities.
First, if the true $\xeff$ distribution is narrower than we can presently resolve with 10 events, then our Gaussian model will generally overestimate the width of the $\xeff$ distribution, thereby \textit{underestimating} the fraction of uninformative events.
Alternatively, if the width of the true $\xeff$ distribution is comparable to or larger than $\Delta$, then the piecewise ``three-bin'' model of \cite{Farr2018} and \cite{O2-Populations} is a poor representation of the true underlying distribution, and will generally \textit{overestimate} the fraction of uninformative events in the central bin.
We may be seeing the effects of one or both of these possibilities, since they are not mutually exclusive.

\vspace{3px}

It is worth noting that \cite{Roulet2019} have also explored a Gaussian model for the binary black hole $\xeff$ distribution, obtaining constraints on $\mu$ consistent with ours when analyzing the GWTC-1 catalogue.
Our measurement of $\sigma^2$, however, qualitatively differs from their results.
Where our posterior permits extremely small $\sigma^2$, \cite{Roulet2019}'s results exclude a vanishingly narrow $\xeff$ distribution.
This distinction could be due to a number of differences between our two analyses.
First, while both analyses adopt a uniform prior in mass ratio, \cite{Roulet2019} additionally assume a uniform prior in chirp mass, while our prior is logarithmically uniform in primary mass.
We have verified, though, that the adoption of a uniform-in-chirp-mass prior does not affect our results.
Second, perhaps more significantly, to generate generate posteriors on $\mu$ and $\sigma^2$, \cite{Roulet2019} rely on analytic likelihood evaluations performed under a number of simplifying assumptions, such as the perfect alignment of all component spins.
On the other hand, we use posterior distributions sampled by MCMC methods.
Finally, \cite{Roulet2019} evaluate the detection efficiency $\xi(\mu,\sigma^2)$ via direct integration, whereas again we take a Monte Carlo approach as described above.
A consequence of our Monte Carlo method is that, numerically speaking, we cannot sample arbitrarily close to $\sigma^2 = 0$.
As $\sigma^2$ approaches zero, the detection efficiency $\xi(\mu,\sigma^2)$ vanishes and the likelihood diverges to infinity.
To prevent this divergence, we impose the stability condition recommended in \cite{Farr2019} -- that the number of synthetic detections contributing to Eq.~\eqref{eqn:selection-sum} be greater than $4 N$ (where $N=10$ is our number of real events).
This condition prevents us from sampling variances below $\sigma^2 \leq 0.03$; even at $\sigma^2=0.03$, though, our posterior remains flat.

\section{Refined Binary Black Hole Spin Measurements}
\label{sec:refined_measurements}

\begin{figure*}
    \centering
    \includegraphics[width=0.9\textwidth]{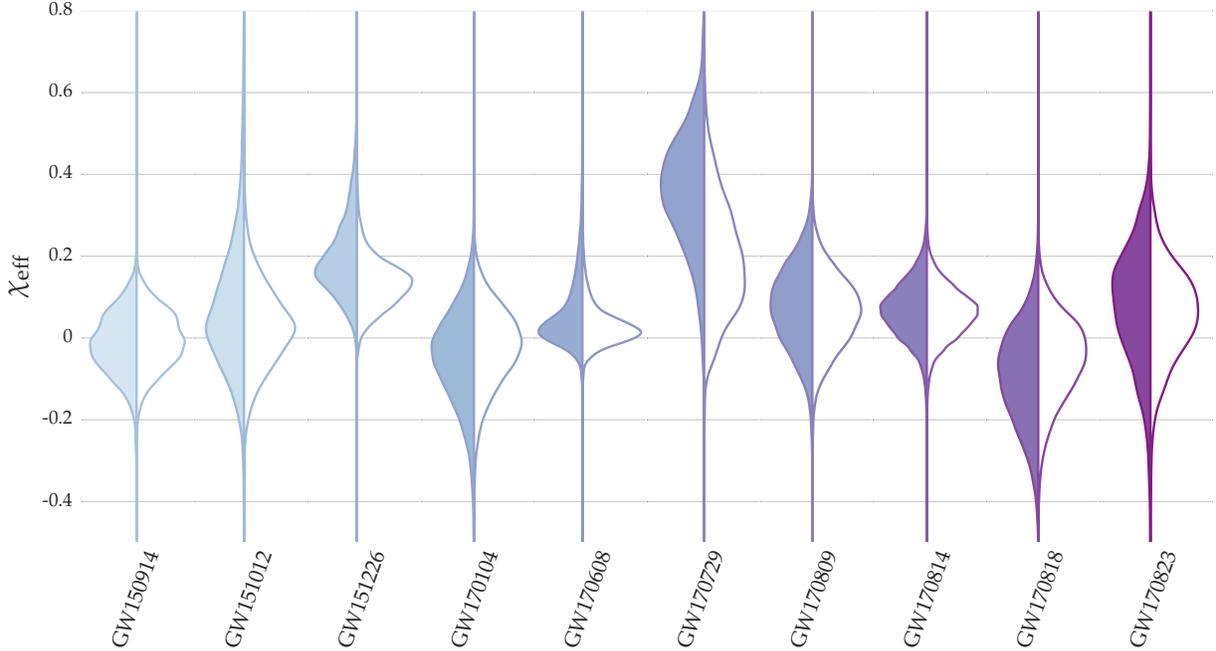}
    \caption{Violin plot showing the posterior distribution on $\xeff$ under the \texttt{LALInference} $\xeff$ prior (filled in) and the population-distribution-informed prior (white) for each binary black hole event.
    The \texttt{LALInference} prior for spin is uniform in spin magnitude and isotropic in spin tilt angle. The colors in this plot are used to distinguish between events, which are plotted in order of detection date.
    Notice that the population-distribution-informed prior brings each posterior distribution closer to $\xeff = 0$.
    This means that when the spin distribution of the binary black hole population is not used to inform parameter estimation, we consistently overestimate the magnitude of $\xeff$.}
    \label{fig:violin_plots}
\end{figure*}

Given what we now know about the population distribution of $\xeff$, we can obtain refined $\xeff$ measurements on each individual binary black hole system.
As described in Sec.~\ref{sec:inference}, the default \texttt{LALInference} priors on component spins are uniform in dimensionless magnitude and isotropic in direction.
Having obtained information about the actual population distribution of $\xeff$, we can replace the uninformative \texttt{LALInference} prior with one that matches the recovered $\xeff$ distribution.
Specifically, we will recompute the posterior $p(\xeff|d)$ on the effective spin of each of the ten binary black hole detections, marginalizing over all possible values of $\mu$ and $\sigma^2$.
We will additionally undo the default \texttt{LALInference} mass and redshift priors, reweighting by our astrophysically-motivated prior in Eq.~\eqref{eqn:pASTRO}.

For brevity, we'll introduce the abbreviation $\lambda=\{\xeff,m_1,m_2,z\}$ for the set of parameters describing an individual compact binary.
Given the complete set $D=\{d_i\}_{i=1}^N$ for all our $N=10$ events, we can form a joint posterior on $\mu$, $\sigma^2$, and the parameters $\lambda$ of some \textit{single} event:
\begin{widetext}
\begin{equation}
\label{eqn:reweighting}
\begin{aligned}
p(\lambda,\mu,\sigma^2 | D)
    &\propto
            p(\lambda \,|\, d) \frac{p(\lambda\,|\,\mu,\sigma^2)}{p_\mathrm{pe}(\lambda)}
            \left(\frac{p(\mu,\sigma^2)}{\xi(\mu, \sigma^2)^N} \, \prod^{N-1}_{i=1}
            \left[\,
                \int d\lambda\,\,
                    p(\lambda \,|\, d_i)
                    \frac{p(\lambda\,|\,\mu,\sigma^2)}{p_\mathrm{pe}(\lambda)}
            \right]\right).
\end{aligned}
\end{equation}
This is simply Eq.~\eqref{eqn:posterior}, absent marginalization over the single event of interest.
Equation~\eqref{eqn:reweighting} can be put into a more convenient form if we multiply and divide by $ p(d|\mu,\sigma^2) = \int d\lambda\,p(\lambda \,|\, d)\frac{p(\lambda\,|\,\mu,\sigma^2)}{p_\mathrm{pe}(\lambda)}$.
This gives
\begin{equation}
\label{eqn:reweighting2}
\begin{aligned}
p(\lambda,\mu,\sigma^2 | D)
    &\propto
             \frac{p(\lambda \,|\, d)}
                {p(d\,|\,\mu,\sigma^2)}
            \frac{p(\lambda\,|\,\mu,\sigma^2)}{p_\mathrm{pe}(\lambda)}
            \left(\frac{p(\mu,\sigma^2)}{\xi(\mu, \sigma^2)^N} \, \prod^{N}_{i=1}
            \left[\,
                \int d\lambda_i\,\,
                    p(\lambda_i \,|\, d_i)
                    \frac{p(\lambda_i\,|\,\mu,\sigma^2)}{p_\mathrm{pe}(\lambda_i)}
            \right]\right) \\
    &\propto
        \frac{p(\lambda \,|\, d)}
                {p(d\,|\,\mu,\sigma^2)}
            \frac{p(\lambda\,|\,\mu,\sigma^2)}{p_\mathrm{pe}(\lambda)}
        p(\mu,\sigma^2\,|\,D).
\end{aligned}
\end{equation}
\end{widetext}
This expression now depends only on the default posterior $p(\lambda|d)$ for the event of interest and the marginal posterior $p(\mu,\sigma^2\,|\,D)$ on the population parameters given \textit{all} $N$ events -- exactly what we computed above in Sect.~\ref{sec:inference}.

Given discrete samples from $p_\mathrm{pe}(\xeff,m_1,m_2,z|d)$ and $p(\mu,\sigma^2|D)$, we can sample Eq.~\eqref{eqn:reweighting} via a two step procedure, in which we (\textit{i}) select a posterior sample $\{\mu_i,\sigma^2_i\}$ from $p(\mu,\sigma^2|D)$, then (\textit{ii}) select a random parameter estimation sample $\lambda_j \equiv \{\chi_\mathrm{eff},m_1,m_2,z\}_j$, subject to the weights
    \begin{equation}
    \label{eqn:reweighting-weights}
    \begin{aligned}
    w_j &=
            \frac{p(\lambda_j|\mu_i,\sigma^2_i)}
            {p_\mathrm{pe}(\lambda_j)} \\
        &=
            \frac{p(\xeffj|\mu_i,\sigma_i^2)\,p(m_{1,j},m_{2,j},z_j)}{p_\mathrm{pe}(\xeffj)\,p_\mathrm{pe}(m_{1,j},m_{2,j},z_j)}.
    \end{aligned}
    \end{equation}
Equation~\eqref{eqn:reweighting2} contains an extra factor, $p(d|\mu,\sigma^2)^{-1}$, that does not appear in the above weights.
This is because, once we condition on the particular values  $\mu_i$ and $\sigma^2_i$, the factor $p(d|\mu,\sigma^2)$ appearing in Eq.~\eqref{eqn:reweighting2} is a constant; we can therefore neglect it from the weights in Eq.~\eqref{eqn:reweighting-weights}.

\vspace{3px}

For each of the ten binary black holes detected in O1 and O2 by Advanced LIGO Virgo, Fig.~\ref{fig:violin_plots} compares the default \texttt{LALInference} $\xeff$ posteriors to our population-informed posteriors.
The \texttt{LALInference} posteriors are shaded, while the population-informed posteriors are in white.
In all cases the population-informed prior serves to shift each posterior distribution towards $\xeff = 0$.
This shift operates in both directions -- posteriors that strongly support positive $\xeff$ are moved downward, while posteriors supporting negative effective spins shift up.  This ``shrinkage'' effect is common to hierarchical models of observations with significant uncertainty \citep{Lieu2017}.
GW170729, for example, excludes $\xeff\leq0$ with 99\% credibility when using uninformative priors.
The population-informed prior, though, noticeably shifts the GW170729's $\xeff$ posterior downward; 8\% of the posterior now extends below $\xeff=0$ and the peak is lowered from $\xeff\approx0.4$ to $0.1$.
The effective spin of GW151226, on the other hand, remains confidently positive, with $\xeff>0$ at 99\% credibility under both the uninformative and population-informed priors.
GW170818, meanwhile, originally shows moderate support for negative effective spins, with 77\% of the posterior at $\xeff<0$.
Under our population-informed prior, 68\% of GW170818's $\xeff$ prior now supports negative values.

\vspace{3px}

A key takeaway from these results is that when a collection of binary black holes are analyzed in isolation under default priors common to such analyses, the magnitude of their effective spins will be consistently overestimated.
When studying future compact binary mergers, it will be essential to evaluate their spins in the context of the broader population, particularly when an event seemingly has confidently positive or negative $\xeff$.
Further updated population-informed posteriors on the parameters of individual gravitational wave events are given in \cite{Fishbach2019_pop} and \cite{Galaudage2019}.

\section{Conclusion}
\label{sec:conclusion}

As we enter deeper into the era of gravitational-wave astronomy, increasingly valuable information will be encoded in the ensemble properties of compact binary mergers.
The distribution of parameters like mass, spin, and redshift across the observable universe's binary black hole population contain essential information about the formation and evolutionary pathways of such systems.
We focus on the population distribution for effective spin, $\xeff$.
This parameter characterizes gravitational-wave strain at leading order, and is thus easily accessible from the data from the small number of LIGO/Virgo events.
Additionally, our focus on this single phenomenological parameter provides a simple and intuitive framework on which future analyses can be built. 
In a complementary paper, for example, \cite{Mohammad} extend our model to investigate correlations between the effective spins and masses of binary black holes.

\vspace{3px}

In this paper, we have explored the distribution of the effective spin parameter $\xeff$ across the ten binary black holes observed by LIGO and Virgo in their O1 and O2 observing run.
We adopted a simple and intuitive model, measuring the mean $\mu$ and variance $\sigma^2$ of effective spins across this ensemble of gravitational-wave events,and rigorously accounting for observational selection effects and degeneracies.
At 95\% credibility, we found $\mu=0.02^{+0.11}_{-0.13}$ and $\sigma^2\leq0.07$.
Notably, $\sigma^2$ is consistent with $0$, meaning that the distribution of BBH effective spins is consistent with a delta function.
As discussed in Sec.~\ref{sec:results}, it is not clear whether this result preferentially supports a dynamical or field binary origin (or even other possibilities) for stellar-mass binary black holes.
Instead, this result likely indicates that binary black holes have small spin magnitudes, far smaller than those inferred in x-ray binaries~\citep{Miller2015}.

\vspace{3px}

Our knowledge of the ensemble distribution of $\xeff$, in turn, allowed us to refine $\xeff$ measurements for individual binary black hole events.
Existing spin measurements were obtained under an uninformative prior on component spin magnitudes and orientations.
Here, we instead used our measurements of $\mu$ and $\sigma^2$ to generate a \textit{population-informed} prior, which we used to reweight existing parameter estimation results for the ten O1 and O2 binary black hole detections.
Using our population-informed prior, the resulting effective spin posteriors are all shifted towards $\xeff = 0$.
In some cases this shift is significant.
For example, GW170729 previously excluded $\xeff=0$ at 99\% credibility.
Under a population-informed prior, however, it shows a $\sim10\%$ probability for zero (or negative) effective spin.
More broadly, our results demonstrate the value of analyzing the ensemble of binary black holes in unison; when treating each binary black hole in isolation, we will consistently overestimate the magnitude (positive or negative) of $\xeff$.

\vspace{3px}


\section*{Acknowledgements}

We would like to thank Christopher Berry, Thomas Dent, Zoheyr Doctor, Maya Fishbach, and others within the LIGO Scientific Collaboration and Virgo Collaboration for constructive comments and helpful conversation.
The Flatiron Institute is supported by the Simons Foundation.
This research has made use of data, software and/or web tools obtained from the Gravitational Wave Open Science Center (https://www.gw-openscience.org), a service of LIGO Laboratory, the LIGO Scientific Collaboration and the Virgo Collaboration. 
LIGO is funded by the U.S. National Science Foundation. 
Virgo is funded by the French Centre National de Recherche Scientifique (CNRS), the Italian Istituto Nazionale della Fisica Nucleare (INFN) and the Dutch Nikhef, with contributions by Polish and Hungarian institutes.

\bibliography{References.bib}

\end{document}